# Risk assessment of muon single-event effects for low-altitude aircraft*

QIU Zijian, LIN Sujie, CUI Yudong, LIU Yihan, YANG Lili*

School of Physics and Astronomy, Sun Yat-Sen University, Zhuhai 519082, China

**Abstract**

**Motivation** With the rapid development of low-altitude economy, the radiation environment safety of low-altitude aircraft such as drones and electric vertical take-off and landing (eVTOL) aircraft has attracted increasing attention. Traditional views hold that the dense lower atmosphere is an effective barrier against cosmic radiation, but the reduced feature sizes of modern integrated circuits (ICs) significantly increase their susceptibility to single-event effects (SEEs). Most traditional studies have focused on the effects of particles such as neutrons and protons, while systematic assessments of the risks induced by muons — the most abundant charged particles at sea level— are still scarce, especially during extreme solar events. Therefore, this study quantitatively evaluates the muon-induced SEE risk for low-altitude aircraft in different regions of China under static cosmic-ray background and ground level enhancements (GLEs), aiming to provide key insights for the operational safety of the next-generation low-altitude aviation platforms.

**Methods** This study employs city-specific atmospheric model to simulate the atmospheric shower processes in different cities within the CORSIKA framework, yielding reliable energy spectra of low-energy muons (10 - 100 MeV) in different regions. This study utilizes electrical simulation data from other research groups to estimate muon-induced SEE cross sections in transistors with different process nodes, covering bulk, FD-SOI, and FinFET processes. Subsequently, by integrating solar energetic particle (SEP) energy spectra related to ground level enhancement (GLE) events, we evaluate the muon-induced SEE risks for systems of different sizes under static conditions (only cosmic-ray injection) and GLE event scenarios.

**Results** Our results indicate that under static conditions, flight control systems (with 1 MB of memory) containing advanced process nodes (⩽ 45nm) and bulk transistors

---

face non-negligible muon-induced SEE risks in all cities in China. In contrast, the systems utilizing FD-SOI transistors can effectively alleviate such risks. For systems with large memory capacity (1 GB), redundancy and other radiation-hardening measures must be taken regardless of the process technology used. Regarding GLE events, this study innovatively introduces the concept of muon hazard levels to assess regional changes in risk. Specifically, during GLEs, the aggravation of muon-induced SEE risks in mid-to-low latitude regions is negligible, whereas such risks significantly increase in high-latitude regions.

# 1 Introduction

For a long time, the academic community has generally believed that the dense lower atmosphere effectively shields cosmic ray radiation. Consequently, research on the radiation hazards of cosmic rays and solar wind particles has primarily focused on the aerospace sector, with limited attention paid to risks in low-altitude flight scenarios. When high-energy particles ($E > 1$ GeV) penetrate the atmosphere, they interact with atmospheric nuclei. This interaction generates a large number of secondary particles, such as muons, protons, electrons, neutrons, and heavy ions, through atmospheric shower processes. The flux of these secondary particles ($E > 1$ MeV) is significantly higher than that of primary cosmic ray particles. Upon striking electronic devices, they deposit charge, leading to functional anomalies or damage. This phenomenon, where a single particle impact causes electronic system anomalies, is known as the single event effect (SEE). It is currently recognized as a major issue affecting the reliability of commercial electronic systems[1].

In recent years, the proliferation of drone logistics and low-altitude sightseeing aircraft has led to an explosive growth in the number of low-altitude vehicles operating below 3 km. Benefiting from advances in integrated circuit technology, these vehicles are typically equipped with sophisticated electronic systems, which effectively reduce volume and energy consumption. However, as transistor process nodes shrink, the minimum deposited charge required to change the transistor state (i.e., the critical charge) also decreases. This allows particles with low energy deposition rates, such as muons, to induce SEE. This phenomenon has been verified by muon beam radiation experiments[2,3], and its impact intensifies as process nodes shrink[2]. Muons are the most abundant charged particles near sea level. Multiple electronic simulation studies indicate that near sea level, muons may become the primary source of SEE when the transistor process node is smaller than 45 nm[4,5]. Therefore, assessing the muon-induced SEE risk in different urban areas is an emerging issue for ensuring the safety of low-altitude traffic operations and promoting the development of the low-altitude economy.

The accuracy of electronic simulation assessments for SEE largely depends on the precise estimation of muon flux, particularly the muon energy spectrum in the MeV range[6,7]. Currently, most related studies employ analytical radiation models based on measured data and Monte Carlo simulations, such as QARM[8,9] and PARAM[10], to derive the differential energy spectrum of muons.

However, these models have two main limitations. First, their predictions for muons cannot be reliably extended to the MeV energy range. Currently, only a few experiments have measured the atmospheric muon differential spectrum below 1 GeV under specific regional and altitude conditions[11-14], and there is only one experimental dataset below 400 MeV[14]. In the MeV range, where experimental data are generally scarce, physical simulations must be relied upon to predict the secondary muon energy spectrum. However, models such as QARM and PARAM do not account for differences in atmospheric models across different regions in their Monte Carlo simulations. Consequently, this introduces systematic biases in the prediction of muon flux in the MeV range.

Second, existing models struggle to effectively assess the increased SEE risk during solar activity periods. During periods of solar activity, extreme solar energetic particle (SEP) events can accelerate particles to energies above GeV. Upon entering the Earth's atmosphere, these particles generate secondary particle showers that extend to the ground. Such events are known as ground level enhancements (GLE). Secondary particles from ground level enhancements (GLEs) were initially detected by ground-based neutron monitors[15]. Subsequently, a few muon detection experiments also observed an increase in muon flux during GLEs[16,17]. However, for muons in the MeV range, no observational or simulation studies have currently provided their flux growth rates during ground level enhancement (GLE) events.

To address these issues, this study introduces the extensive air shower simulation package CORSIKA[18]. By systematically simulating atmospheric shower processes under different atmospheric conditions and solar activity scenarios, we comprehensively estimate the muon energy spectrum in the MeV range. This provides reliable data support for assessing muon-induced SEE risks.

This study considers two scenarios: solar quiet periods and solar active periods. We configured different atmospheric models for different cities and conducted complete Monte Carlo simulations to assess the muon-induced SEE risk faced by low-altitude vehicles in the low-altitude environments of various urban areas. The structure of this paper is as follows: Section 2 details the Monte Carlo simulation method for low-altitude muon flux and its validation. Section 3 combines electronic models to assess muon-induced SEE risks in different cities and scenarios. Section 4 discusses the uncertainties and limitations of the study. Section 5 summarizes the paper.

# 2 Simulation of Low-Altitude Muon Flux

## 2.1 Simulation Setup

In this work, the four-dimensional simulation package CORSIKA [18] (version 7.7500) for extensive air showers is employed to obtain the differential energy spectrum of muons in low-altitude environments. CORSIKA provides a variety of hadronic interaction models. In this study, the combined model of EPOS-LHC [19] and UrQMD [20, 21] is adopted for the simulation. The EPOS-LHC model incorporates constraints from LHC collision experimental data and is widely used in the simulation of high-energy cosmic-ray showers [22]. However, it is not applicable at low energies (<80 GeV). Therefore, the UrQMD model is used as a substitute for simulating hadronic interactions at the low-energy end. The transition threshold between the high-energy and low-energy models is set to the default value, i.e., $E_{\mathrm{Lab}}$=80 GeV/$n$ (where $n$ is the number of nucleons in the projectile particle).

After determining the interaction models, the uncertainty in the simulation results mainly stems from the primary particle injection spectrum and the atmospheric model. Regarding atmospheric models for different regions, this study selects several typical cities in China with different climate types as samples: Shenzhen (subtropical), Qingdao (mid-latitude coastal), Hohhot (mid-latitude inland), and Lhasa (plateau). The North Pole is also selected as a high-radiation area for comparison. The relevant latitude, longitude, and geomagnetic field information are listed in Table A1. The geomagnetic field information is calculated using the 14th generation International Geomagnetic Reference Field (IGRF)[23]. It is important to note that due to the different average elevations of each city (as shown in Table A1), the actual height above ground at the same altitude varies significantly across cities. Therefore, the low-altitude environment defined in this study refers to the region within 3000 m above the local ground surface.

## 2.2 Primary Particle Spectrum

Ground-level muons all originate from the atmospheric shower processes of high-energy particles. This study defines the differential energy spectrum of particles incident on the atmosphere as the primary particle spectrum. Minimizing the uncertainty of the primary particle spectrum is key to improving simulation accuracy. This section describes the selection of primary particle spectra for two scenarios: static (only high-energy cosmic ray injection) and solar active periods (SEP injection superimposed on the static baseline).

Although cosmic rays contain various high-energy particles, this study considers only the five elements with the largest contributions to save computational resources: hydrogen (H), helium (He), carbon (C), oxygen (O), and iron (Fe). For SEP injection, only the main component, protons, is considered. For different atmospheric model configurations, the simulated quantities of each primary particle in different energy ranges are listed in Table A2. Within each energy range, particle energies are sampled according to a power-law spectrum (spectral index of -1.5). The target energy spectrum is simulated by assigning weights to the events.

### 2.2.1 High-Energy Cosmic Ray Spectrum

The Alpha Magnetic Spectrometer 2 (AMS-02) on the International Space Station has precisely measured the spectra of various cosmic ray particles in the GeV-TeV energy range[24]. Although the solar magnetic field modulates cosmic rays and this effect varies with time, the impact of solar modulation is negligible for high-energy cosmic rays with sufficient energy to generate secondary particle showers. Therefore, this study directly adopts the observational results of AMS-02 and their uncertainties ($1\sigma$) as the primary cosmic ray spectrum in the GeV-TeV range.

For the energy range above TeV, both space satellite and ground-based observation experiments exhibit significant uncertainties and systematic biases, making them unsuitable for direct use as primary injection spectra. Gassier et al.[25] considered all available high-energy observation experiments in their 2013 work. Based on simple physical assumptions, they proposed an analytical energy spectrum model for high-energy cosmic rays. This model can effectively reduce the systematic bias between single experiments and the true energy spectrum. Therefore, this study adopts the Gassier model as the primary injection spectrum at the high-energy end (Figure 1). Since the Gassier model does not provide uncertainty estimates, we assume a relative error of 30% ($1\sigma$).

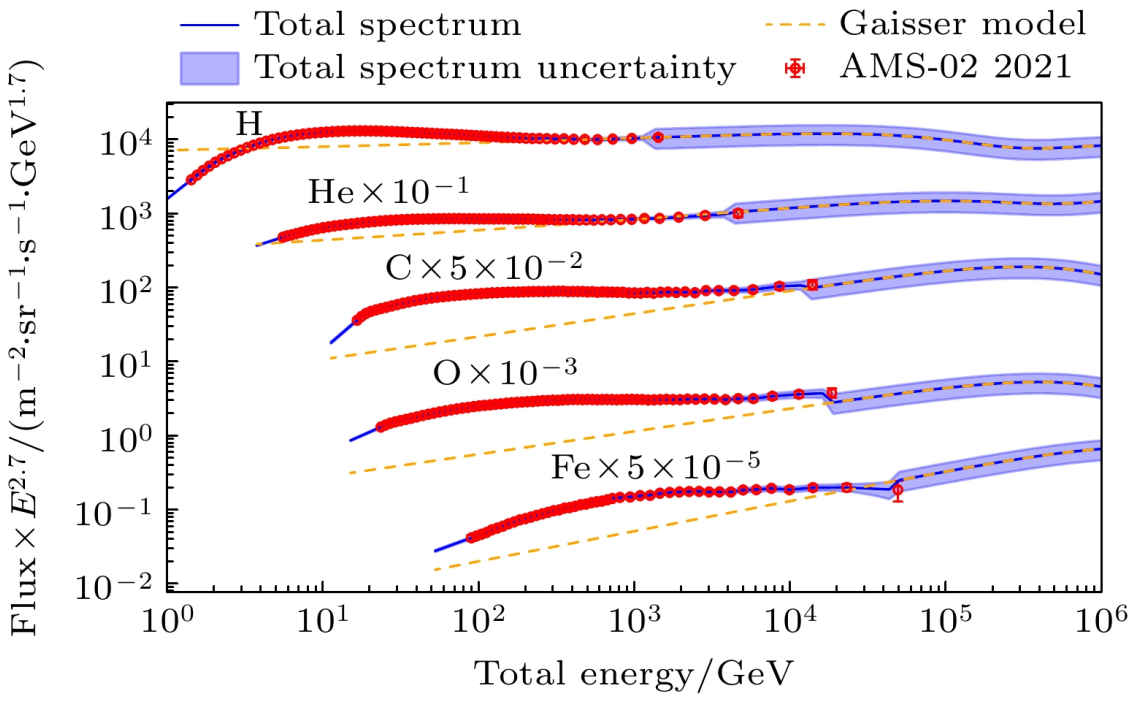

**Fig. 1 Primary cosmic ray spectra. Blue line/shade: primary injection spectrum & uncertainty (this work); red points: AMS-02 data; orange dashed line: Gassier model. All non-proton**

**spectra are scaled.**

### 2.2.2 SEP Energy Spectra

Although GLE events have been observed for over 50 years, the mechanism by which solar activity accelerates protons to the GeV energy range remains controversial. Therefore, this study employs injection simulations based on observational SEP energy spectra. Satellite instruments directly observe SEP spectra in the MeV range, whereas spectra above GeV energies rely primarily on indirect observations from ground-based neutron monitors. Tylka and Dietrich[26] developed a method in 2009 to analyze data from the global neutron monitor network. By combining satellite and neutron monitor data, they derived the integral proton energy spectra for 53 of the 66 GLE events that occurred between 1956 and 2009. The results indicate that the integral rigidity spectrum of SEPs injected into the atmosphere during GLE events can be described by a double power-law function (namely, the Band function[27]):

$$J(>R)=\begin{cases} J_0\left(\frac{R}{1\mathrm{GV}}\right)^{-\gamma_1}\exp\left(-\frac{R}{R_0}\right), & R<R_1, \\ J_0\left(\frac{R_1}{1\mathrm{GV}}\right)^{-\gamma_1}\exp\left(-\frac{R_1}{R_0}\right)\left(\frac{R}{R_1}\right)^{-\gamma_2}, & R\geqslant R_1, \end{cases} \tag{1}$$

Here, $J(>R)$ denotes the total proton flux with rigidity greater than $R$; $R_1=(\gamma_2-\gamma_1)R_0$; and $J_0, \gamma_1, \gamma_2, R_0$ represent the normalization coefficient of the integral spectrum, the spectral indices corresponding to the two power-law segments, and the rigidity at the power-law break, respectively.

Raukunen et al.[28] employed the same method in their 2018 study to fit the integral proton spectrum parameters for 59 GLE events occurring between 1956 and 2012. This study adopts the results from Raukunen et al. as the SEP injection spectra during GLE events, assuming that particles inject uniformly within an integration time window of 1800 s. It is important to note that the azimuthal dependence of the SEP energy spectrum and the maximum energy cutoff remain unresolved issues. We assume an SEP energy cutoff of 100 GeV and simplify the angular distribution into two scenarios: 1) isotropic injection; and 2) extreme anisotropic injection (where particles concentrate from directions with a zenith angle $<20°$). The second scenario serves as an upper-bound estimate for risk assessment.

### 2.2.3 Geomagnetic Shielding

The Earth's magnetic field deflects and shields charged particles, such as SEPs and cosmic rays, via the Lorentz force, thereby reducing the particle flux reaching near-Earth space. The geomagnetic shielding effect varies by location and is typically quantified by the cut-off rigidity ($R_c$): only charged particles with rigidity greater than $R_c$ can penetrate the magnetic shielding. Given that particles inject from an altitude of approximately 100 km in CORSIKA simulations, this study employs the IGRF to calculate the cut-off rigidity at 100 km for each study site (Table A1)[29] to quantify the intensity of geomagnetic shielding.

## 2.3 Atmospheric Model

This study focuses on the muon energy spectrum below 1 GeV, where regional differences in atmospheric models cannot be ignored. Processes such as particle collisions and energy loss in atmospheric showers depend primarily on gas density; therefore, the main influencing factor is the atmospheric density distribution function (i.e., the variation of density with altitude). We utilize the gdastool in the CORSIKA software, based on the Global Data Assimilation System (GDAS) database, to generate atmospheric density distribution functions for different regions.

This distribution exhibits short-term weather fluctuations and long-term seasonal variations. To evaluate the uncertainty introduced by the atmospheric model, we uniformly select 300 time points throughout 2024 for each city to generate the corresponding atmospheric density distribution functions. To conserve the substantial computational resources required for full simulations, we select two representative atmospheric density distribution functions from these points for complete simulation, thereby estimating the uncertainty associated with the atmospheric density model. To this end, we introduce the physical quantity $H$ in Equation (2) to characterize the overall “thickness” of the atmosphere, anticipating a significant correlation with the low-energy muon integral flux (10—100 MeV).

$$H = \frac{1}{5 \times 10^4} \sum_{i=0}^{i<5\times10^4} \rho\,(h_0 + i \times 1\mathrm{m}), \qquad (2)$$

where $\rho(h)$ represents the atmospheric density at altitude $h$, and $h_0$ denotes the altitude of the observation site. During the simulation, the low-energy muon flux is primarily influenced by the lower atmosphere, including the stratosphere and troposphere ($h < 50000$ m), due to the limited penetration capability of low-energy muons. Therefore, summing the densities at multiple points within the lower atmosphere effectively evaluates the overall atmospheric "thickness". To validate the effectiveness of $H$, we conducted small-scale simulations for each atmospheric model (injecting $3 \times 10^5$ protons at 100 GeV, with $h_0 = 50\mathrm{m}$ and $R_c = 9.71\mathrm{GV}$) to obtain the integral flux of low-energy muons corresponding to different $H$ values, as shown in Figure 2. For comparison, we also considered the scenario where the atmospheric density at the observation site, $\rho(h_0)$, was directly used as the "thickness" parameter $H$ (blue dots in Figure 2). The distribution of $\rho(h_0)$ against the low-energy muon integral flux exhibits significant dispersion, indicating that it cannot serve as a reliable "thickness" parameter for quantifying the low-energy muon integral flux.

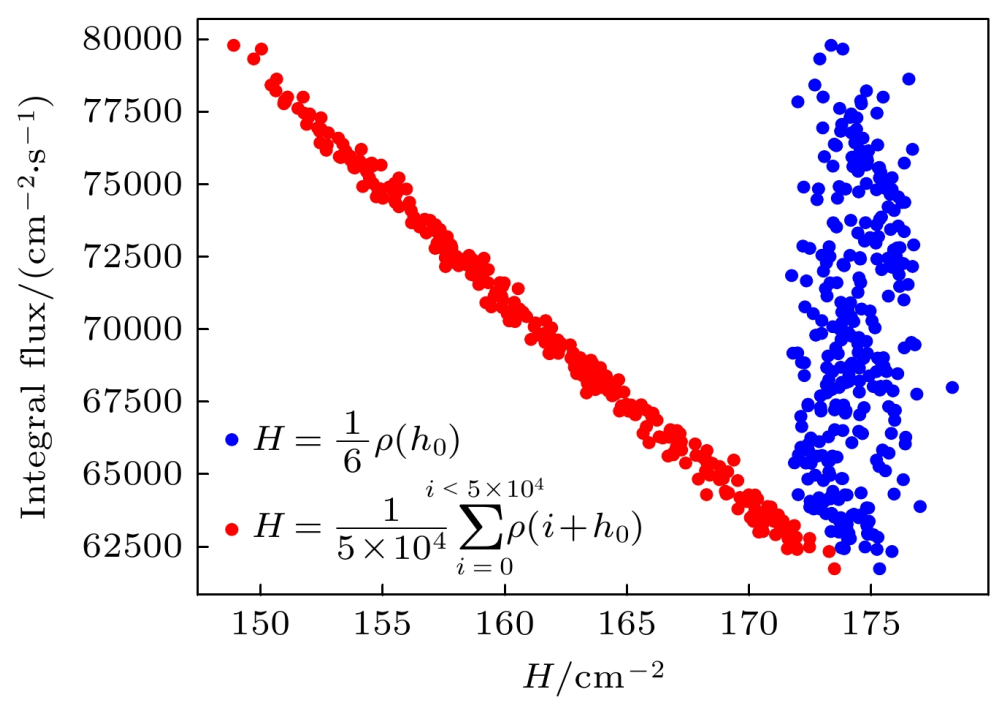

**Fig. 2 Integrated flux of low-energy muons (10–100 MeV) as a function of the "thickness" parameter H. Red points represent the 'thickness' parameter defined by Eq. 2, while blue points represent the parameter $H = \frac{1}{6}\rho(h_0)$.**

Figure 2 reveals a significant monotonic relationship between the low-energy muon integral flux and the $H$ value. Consequently, we rank the atmospheric models by their $H$ values. We select the models corresponding to the 16th and 84th percentiles for comprehensive simulations. The resulting outputs serve as estimates for the $1\sigma$ upper and lower bounds of uncertainty in the low-energy muon integral flux.

Comprehensive simulations yield the low-energy muon integral flux for various cities under static conditions. To visually demonstrate the impact of different atmospheric models on simulation results, this study employs the 1976 US Standard Atmosphere model as input for a complete simulation. We compare these results with simulations for other cities that utilize their respective local atmospheric models (Figure 3). The observed discrepancies indicate the necessity of using region-specific atmospheric models for simulations. Furthermore, regions other than Shenzhen exhibit approximately 15% uncertainty due to variations in atmospheric models. In contrast, Shenzhen is located in a subtropical zone with strong year-round convection. The seasonal variations in its stratosphere and troposphere are minimal, leading to reduced uncertainty from atmospheric modeling.

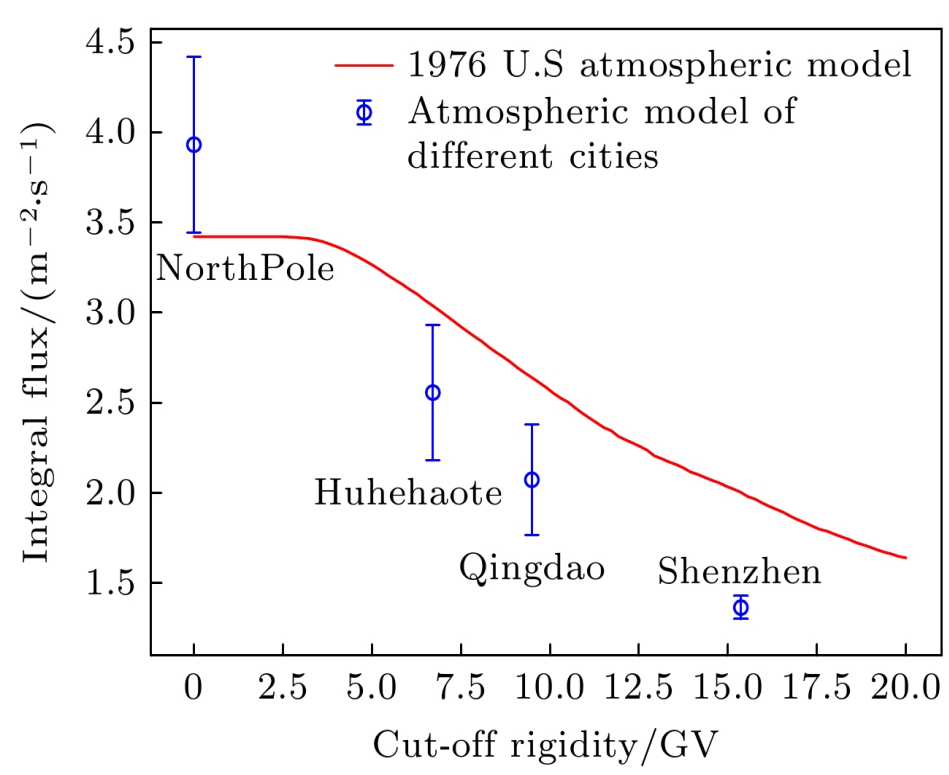

**Fig. 3 Integral flux of low-energy muons (10–100 MeV) as a function of the geomagnetic cutoff rigidity ($R_c$). The red line shows the simulation results using the 1976 U.S. Standard Atmosphere model, while the blue dots represent the results from simulations with the corresponding urban atmospheric model.**

## 2.4 Comparison between Simulation and Experiment

To validate the simulation accuracy, this study compares the results with muon measurements from the BESS spectrometer (Balloon-borne Experiment with a Superconducting Spectrometer) (Figure 4). The BESS spectrometer detects muons with energies down to 0.5 GeV. It has conducted high-precision measurements at three locations with different altitudes and geomagnetic cutoff rigidities: Lynn Lake[11], Mount Norikura[12], and Tsukuba[13]. These measurements provide strict constraints on the muon energy spectrum below 10 GeV. Table A1 lists the details of these three measurement sites. This study applies the corresponding measurement conditions to the simulation to reproduce the observational data.

The atmospheric model represents the primary source of uncertainty. Since the GDAS database lacks observational data from the experimental period, the atmospheric density distribution function during the BESS observations cannot be directly obtained. Therefore, this study selects atmospheric models from 2014 density distribution samples that meet the following criteria as references for the simulation:

$$|\rho(h_0) - \rho_0| \leqslant \sigma_\rho, \quad (3)$$

where $\rho(h_0)$ denotes the atmospheric density at the measurement site within the sample; $\rho_0$ represents the atmospheric density recorded by BESS, and $\sigma_\rho$ is its standard deviation. For the Mount Asama and Tsukuba experiments, the relevant parameters are provided in the corresponding literature. The Lynn Lake experiment, however, comprises three independent measurements, each with distinct atmospheric densities. Therefore, we calculated the weighted average of atmospheric density based on the muon count for each measurement (yielding $\rho_0 \approx 994\mathrm{g/cm^2}$) and adopted a standard deviation of $\sigma_\rho = 5\mathrm{g/cm^2}$. We then executed the selection process described in Eq. (3) using these values.

Furthermore, we sorted the selected reference samples according to the "thickness" defined in Eq. (2). We performed full simulations using atmospheric density models corresponding to the upper and lower $1\sigma$ limits to determine the $1\sigma$ uncertainty range for the simulated muon flux. Figure 4 presents a comparison between the simulated muon intensity and the BESS experimental measurements.

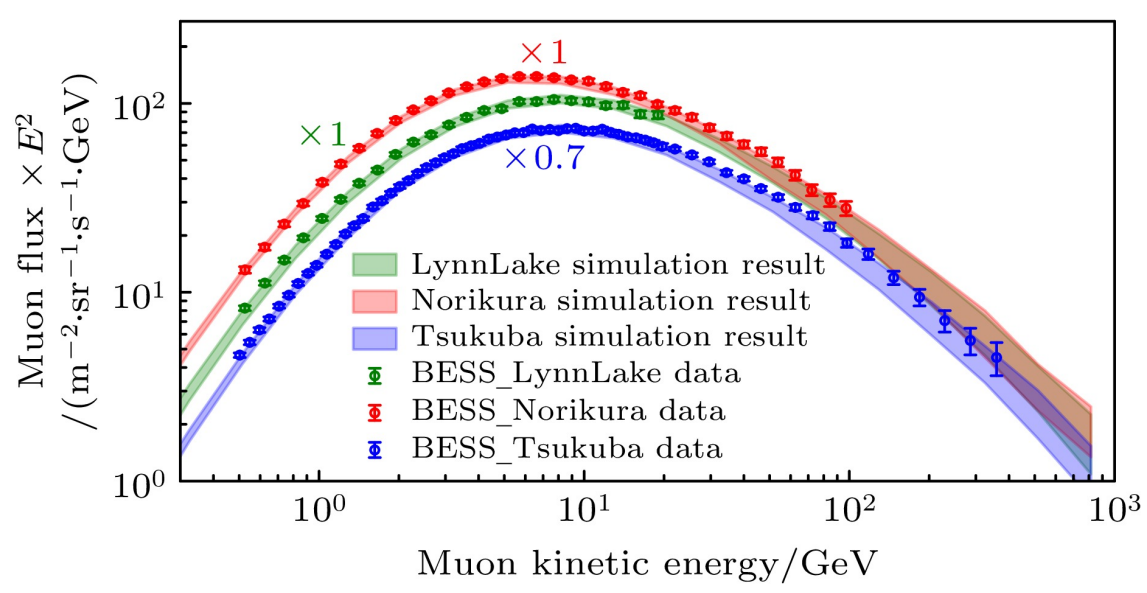

**Fig. 4 Comparison of measured and simulated muon energy spectra from different geographical locations. The shaded areas represent the simulation results with $1\sigma$ uncertainty, while the data points are the measurements from the BESS experiment. Different colors correspond to different locations: green for Lynn Lake, red for Norikura, and blue for Tsukuba.**

The simulation results in this study accurately reproduce the muon detection outcomes of the BESS experiment. Differences in muon energy spectra across various regions are primarily influenced by cutoff rigidity, altitude, and atmospheric models. The uncertainty in muon flux originates from atmospheric models and the primary particle spectrum: the former dominates at the low-energy end, while the latter governs the high-energy end above 10 GeV.

# 3 Muon SEE Risk Assessment

## 3.1 Single-Event Upset Cross-Section

Muons exhibit a low energy deposition rate in media. Consequently, the single-event effects (SEE) they induce are typically non-destructive. The most common occurrence is the single-event upset (SEU), where the state of a memory or logic unit inadvertently flips from "0" to "1" or vice versa. This state can be restored following a reboot or correction. SEU events are generally detectable and correctable via error-correcting code (ECC) techniques, resulting in limited practical impact. However, as transistor feature sizes shrink, SEU events are more likely to evolve into multiple-cell upsets (MCU), where multiple adjacent memory or logic units simultaneously undergo state flips. Even with ECC, MCUs can still cause device failures. Therefore, MCU is a critical consideration in muon SEE risk assessment.

Comprehensive atmospheric shower simulations yield low-altitude muon energy spectra for different urban areas. To assess the corresponding muon SEE risk, it is necessary to obtain the cross-section for muon-induced MCU events (which corresponds to the probability of a muon causing an MCU). Currently, only a few irradiation experiments have measured the muon MCU cross-section for specific static random-access memory (SRAM) devices [30,31]. These data are insufficient to support systematic research on muon SEE risks for electronic devices with various feature sizes and manufacturing processes.

This study references the work of Hubert et al. [4] from 2018. Based on the Multi-Scale Single-Event Phenomena Prediction Platform (MUSCA SEP3) [32], that study used electronic simulations to determine MCU event rates in nanoscale chips under specific muon fluxes. The results cover three mainstream transistor technologies (bulk silicon, FinFET, and fully depleted silicon-on-insulator [FD-SOI]) and various feature sizes (14—45 nm). By integrating this work, we can comprehensively evaluate the muon SEE risks faced by different transistor technologies and feature sizes in low-altitude flight scenarios.

The work by Hubert et al. provided only MCU event rates. We must estimate the muon MCU cross-section from these data. Charged particles primarily deposit energy through ionization in media, with most energy deposition concentrated at the end of their range, forming a "Bragg peak." The energy loss rate is low before the peak, and energy is rapidly depleted afterward. For muons, sufficient energy to induce SEE is deposited only when the Bragg peak is located near the sensitive volume of the device. This scenario primarily corresponds to low-energy muons of approximately 10 MeV[4]. Considering the shielding effects of chip packaging and aircraft hulls (the range of 10 MeV muons in aluminum is approximately 0.4 cm, whereas this distance shortens to about 1 μm for 1 MeV muons) and combining these with muon radiation experimental results [6,7], we assume that muon SEE is primarily concentrated in the 10—100 MeV energy range. Accordingly, we can estimate the cross-section for MCU events caused by muons incident on nanoscale chips from the work of Hubert et al.:

$$\sigma = \frac{R}{\int_{10\mathrm{MeV}}^{100\mathrm{MeV}} f_{\mu}(E)\mathrm{d}E}, \qquad (4)$$

here $\sigma$ denotes the collision cross-section for MCU events induced by low-energy muons, $f_{\mu}(E)$ represents the muon differential flux, and $R$ indicates the MCU event rate derived from electronic simulations (unit: FIT/Mbit, where 1 FIT = $10^{-9}\mathrm{h}^{-1}$, corresponding to one event per $10^{9}\mathrm{h}$). The calculated MCU collision cross-sections are subsequently used to compute error rates, as illustrated in Figure 5. To validate the reliability of these electronic simulations, the figure also presents recent muon irradiation experimental results [30,31] for comparison (assuming a positive-to-negative muon flux ratio of 1:1). It is important to note that muon irradiation experiments suggest $\mu^{-}$ are more likely to induce single-event effects (SEE). However, given the significant uncertainty associated with this finding, this study does not differentiate between positive and negative muons; thus, $f_{\mu}(E)$ represents the total flux of both positive and negative muons.

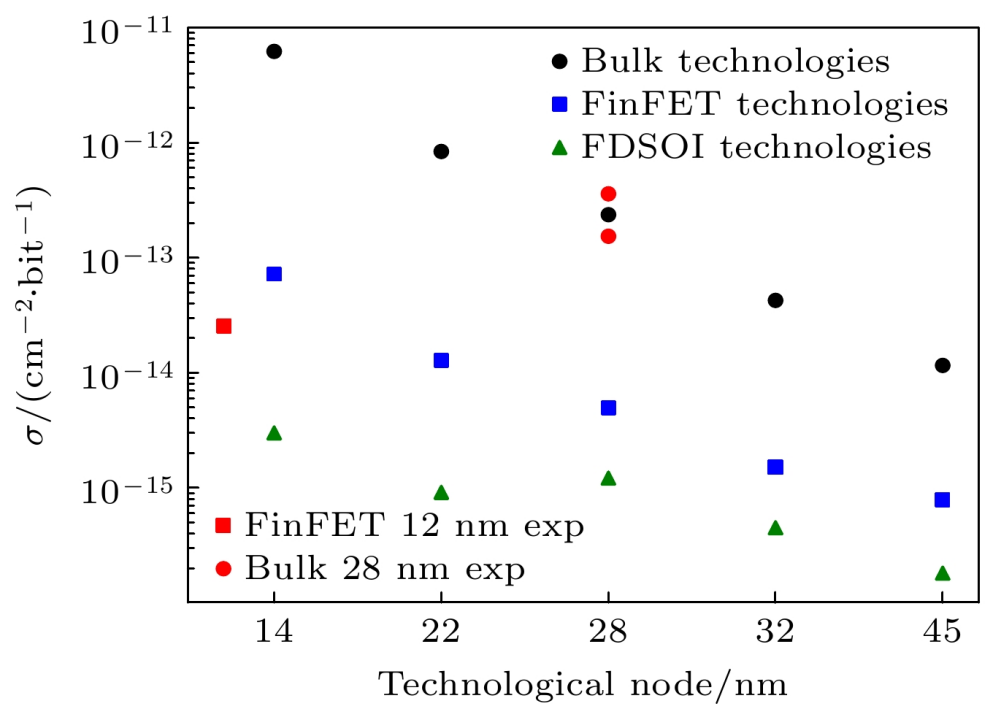

**Fig. 5 MCU cross-section for various transistor technologies and process nodes. Symbols: black dots for Bulk, blue squares for FinFET, and green triangles for FD-SOI. Red symbols denote experimental data: squares for 12 nm FinFET[31] and dots for 28 nm Bulk[30,31]. Note: The 14 nm FD-SOI data point is an extrapolated value.**

## 3.2 Static Low-Altitude Muon SEE Risk

In electronic system safety assessments, the soft error rate (the probability of recoverable device failures per unit time) serves as a key metric for system reliability, with single-event effects (SEE) constituting the primary source of soft errors. Consequently, existing soft error rate standards can assess the muon-induced SEE risk for low-altitude aircraft. As discussed in Section 3.1, only multiple-cell upset (MCU) events cause functional anomalies in practical systems employing error-correcting codes. Therefore, this study uses the MCU event rate instead of the soft error rate as the evaluation metric.

Currently, no soft error rate standards specifically address low-altitude aircraft. Given that flight control systems in these aircraft execute extensive automatic controls during operations such as obstacle avoidance and route navigation, their functional safety requirements and radiation environments closely resemble those of automotive autonomous driving. Thus, this study references the international standard for automotive functional safety, ISO 26262[33]. This standard defines Automotive Safety Integrity Levels (ASIL) and their requirements, explicitly mandating that electronic systems maintain sufficiently low soft error rates. Specifically, the highest safety level, ASIL-D, requires a soft error rate $\leqslant 10$ FIT.

Here, we calculate the upper limit of the allowable low-energy muon integral flux for chips of different process nodes and technologies to meet the ASIL-D level ($\leqslant 10$ FIT), as shown in Figure 6. We consider two typical system memory capacities: 1 MB (corresponding to the typical SRAM capacity of current low-altitude aircraft flight control systems) and 1 GB (corresponding to the typical capacity of airborne computers).

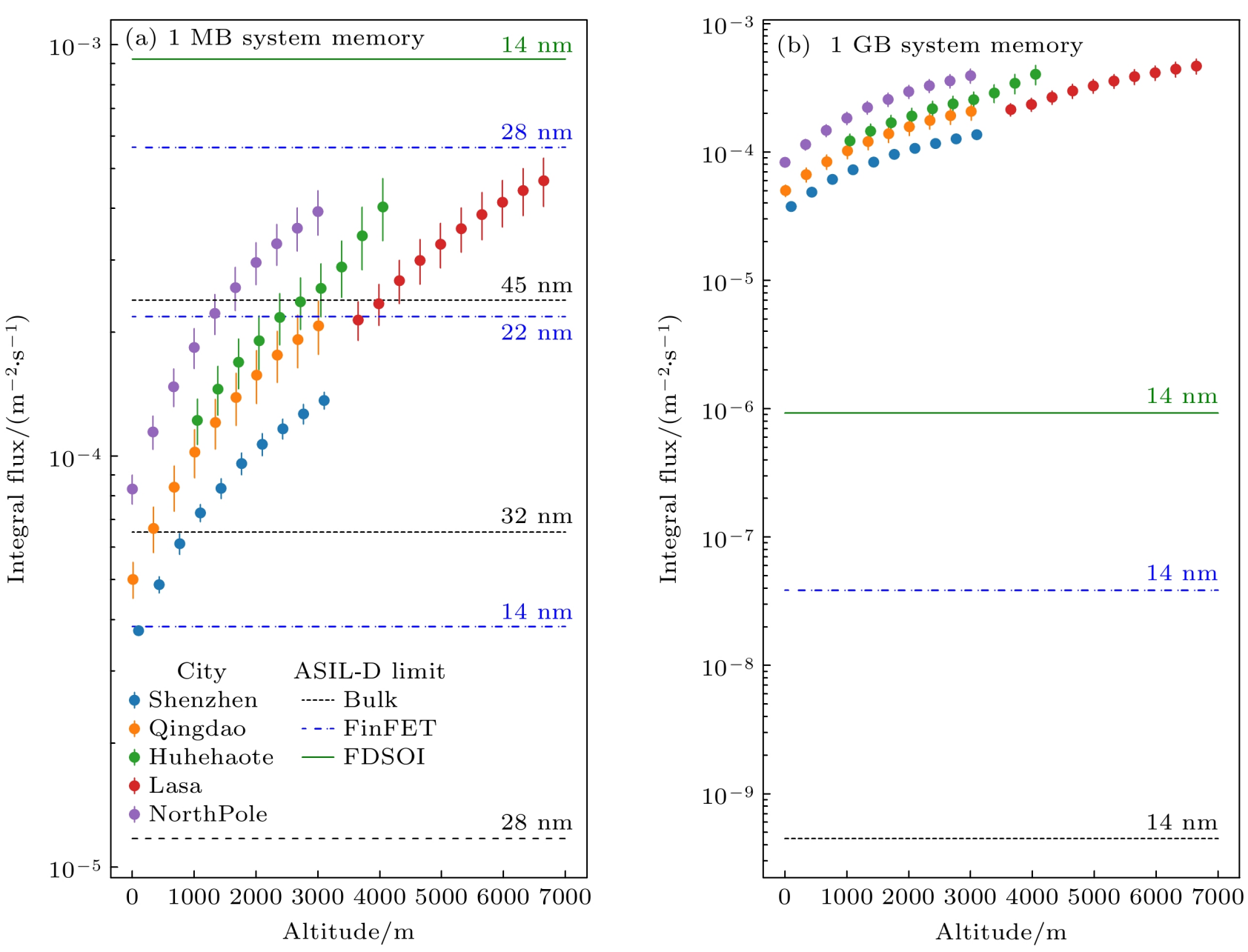

**Fig. 6 Variation of the simulated low-energy muon (10–100 MeV) integral flux with altitude for different cities, in comparison with the ASIL-D permissible flux limits. (Left panel: 1 MB system memory; Right panel: 1 GB system memory). The safe flux limits for FD-SOI (green solid line), FinFET (blue dashed line), and Bulk (black dotted line) transistors are shown, with their corresponding process nodes labeled.**

Figure 6 illustrates the variation in low-energy muon integral flux (10-100 MeV) with altitude across different cities under static conditions. The black dashed lines represent the upper flux limits required to meet the ASIL-D safety level ($\leqslant$ 10FIT) for various combinations of memory capacity and chip technology nodes. This visualization allows for an intuitive assessment of muon-induced single-event effect (SEE) risks across different regions, chip technologies, and manufacturing processes.

For flight control systems (1 MB memory): If advanced bulk technology nodes of 45 nm or smaller are employed, significant muon SEE risks persist across all regions. Specifically, in high-latitude (North Pole) and high-altitude (Lhasa) areas, the muon flux exceeds the ASIL-D limit for 45 nm bulk technology. Consequently, larger technology nodes or additional protective measures are necessary. FinFET and FD-SOI technologies exhibit significantly stronger radiation hardness. Notably, FD-SOI technology meets ASIL-D requirements even at the advanced 14 nm node in high-latitude regions. For FinFET technology, technology nodes of 28 nm or larger should be considered.

For large-memory systems such as airborne computers (1 GB memory): Even when using FD-SOI technology, which offers the strongest radiation hardness, the event rate for microcontroller units (MCUs) at the 45 nm node far exceeds the ASIL-D limit. Therefore, the use of small-technology-node chips in critical systems should be avoided, or additional hardening measures, such as triple modular redundancy, must be implemented.

### 3.3 Low-altitude muon SEE risk during GLE events

Currently, ground-level enhancement (GLE) events are primarily observed via satellite and ground-based neutron monitoring networks. From 1956 to 2012, a total of 67 GLE events were recorded. Among these, the GLE event on February 23, 1956, was the largest in the past century. To verify whether the simulation accurately reproduces GLE events, Figure 7 compares the neutron flux increase rates recorded by various neutron monitoring stations during the 1956 GLE event[34] with the atmospheric shower simulation results. The integration time was kept consistent, the 1976 US Standard Atmosphere model was used for all simulations, and the geomagnetic cutoff rigidity for each station is listed in Table 1 of Reference [34]. The comparison indicates that the simulation basically reproduces the observations from neutron monitors. Regarding the GLE event on December 13, 2006, muon detection experiments observed a 0.61% increase in muon flux during the event (integration time: 10 min)[17]. Based on the location information provided in that study, our simulation yielded a muon flux increase rate of 0.78% (isotropic injection), which aligns with the observational data.

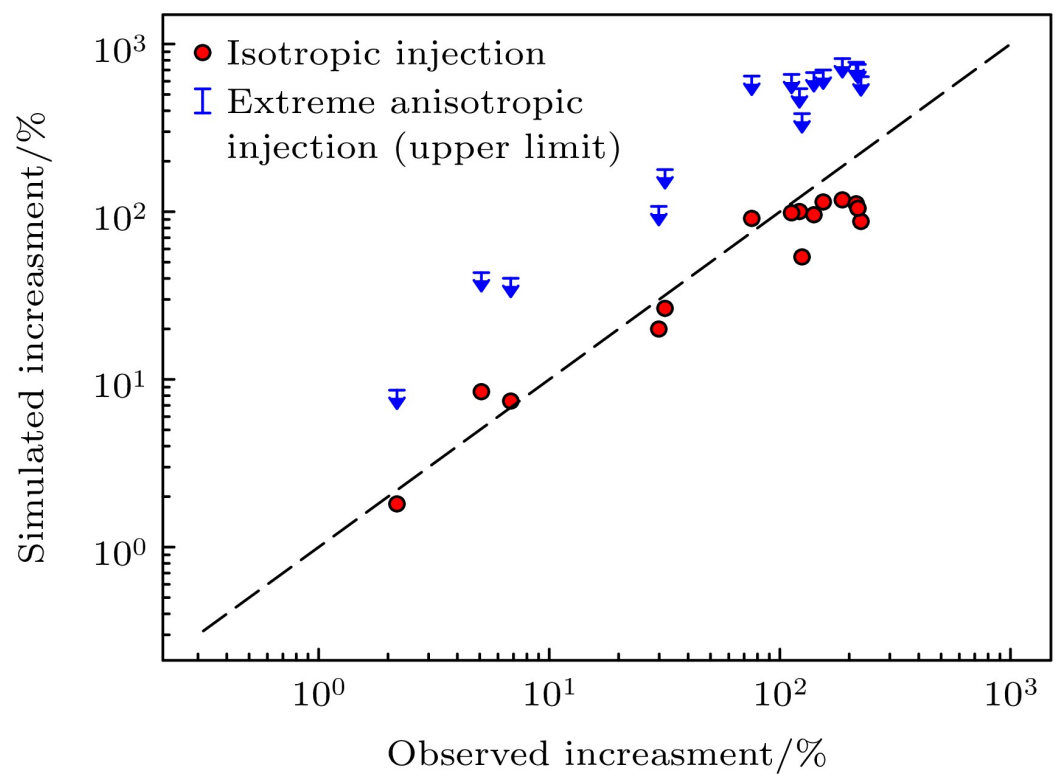

**Fig. 7 Comparison of the observed neutron monitor count rate increase with simulation results for the 1956 GLE event. Red dots indicate the simulation for an isotropic injection, while blue dots represent the simulation for an extreme anisotropic injection, which is provided as an upper-limit estimate. The dashed line shows a perfect match between simulation and observation for comparison.**

In contrast to the routine monitoring of GLE events by neutrons, experimental observations of muon flux increases are rare. This rarity arises because the SEP energy spectrum is very steep, resulting in a scarcity of high-energy particles above 10 GeV. Consequently, during GLE events, the flux of low-energy muons increases significantly, whereas the flux of muons above 10 GeV remains essentially unchanged. This phenomenon is particularly pronounced at high latitudes (Figure 8). The surge in low-energy muon flux during GLE events exacerbates the risk of muon-induced single-event effects (SEE) in electronic devices. Furthermore, the significant regional variations indicate that the level of risk differs across locations.

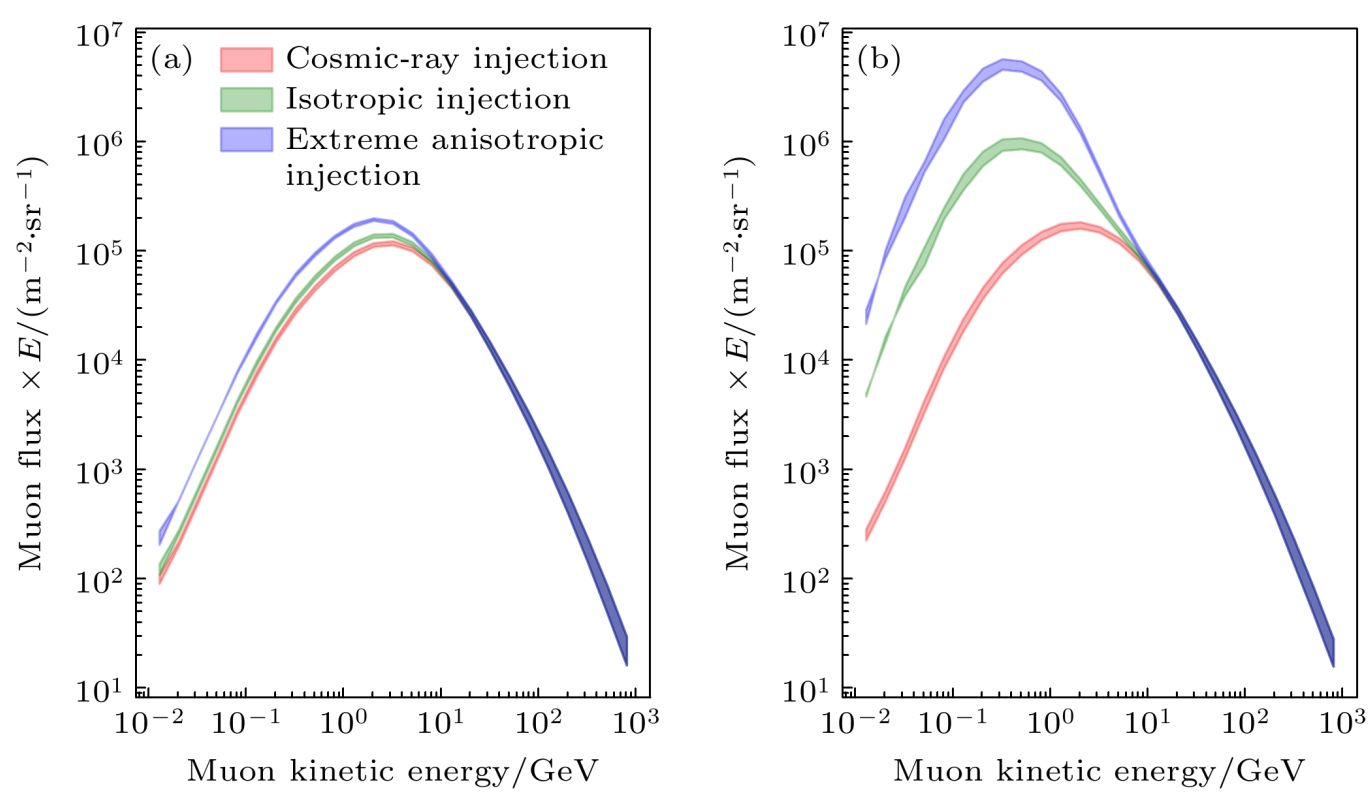

**Fig. 8 Muon energy spectra at an altitude of 3000 m for different regions during the 1956 GLE event, compared with the steady-state background from only cosmic-ray injection: (a) Shenzhen; (b) the North Pole.**

The growth rates of low-altitude muon flux vary significantly across different GLE events. To quantitatively assess the muon SEE risk during GLE events, this study selects the muon integral flux (10-100 MeV) $I_0 = 4 \times 10^{-5} \mathrm{m}^{-2} \cdot \mathrm{s}^{-1}$ as the baseline value. We define the muon hazard level of a GLE event based on the ratio of the muon integral flux $I_{\mathrm{GLE}}$ generated by SEP at an altitude of 3000 m to $I_0$:

1) $I_{\mathrm{GLE}}/I_0 < 1$, defined as a Level I event;
2) $1 \leqslant I_{\mathrm{GLE}}/I_0 < 10$, defined as a Level II event;
3) $10 \leqslant I_{\mathrm{GLE}}/I_0 < 100$, defined as a Level III event;
4) $I_{\mathrm{GLE}}/I_0 \geqslant 100$, defined as a Level IV event.

The baseline value $I_0$ holds dual significance. First, for a system memory of 1 MB, $I_0$ corresponds to the ASIL-D flux upper limit for 14 nm FinFET transistors. If $I_{\mathrm{GLE}} < I_0$, the muon SEE risk generated by the GLE event poses no threat to radiation-hardened FinFET and FDSOI transistors. Second, $I_0$ corresponds to the static muon integral flux at ground level in low-latitude regions (such as Shenzhen). If $I_{\mathrm{GLE}} < I_0$, the muon integral flux generated by the GLE event remains lower than the low-altitude static flux in any region. Consequently, it does not cause a surge in regional muon SEE risk compared to static conditions. Therefore, the muon SEE risk from Level I GLE events is negligible. During Level II events, the muon SEE risk increases noticeably. For Level III events, the risk rises significantly, warranting caution for low-altitude flight missions. Level IV extreme events may cause muon SEE risks to exceed conventional design margins, posing severe threats to electronic systems.

Figure 9 presents the counts of muon hazard levels for 59 GLE events across different cities from 1956 to 2012. This visualization allows for an intuitive assessment of the exacerbated muon SEE risk caused by GLE events in various regions. The data indicate that the muon hazard level of GLE events depends significantly on the geomagnetic cutoff rigidity.

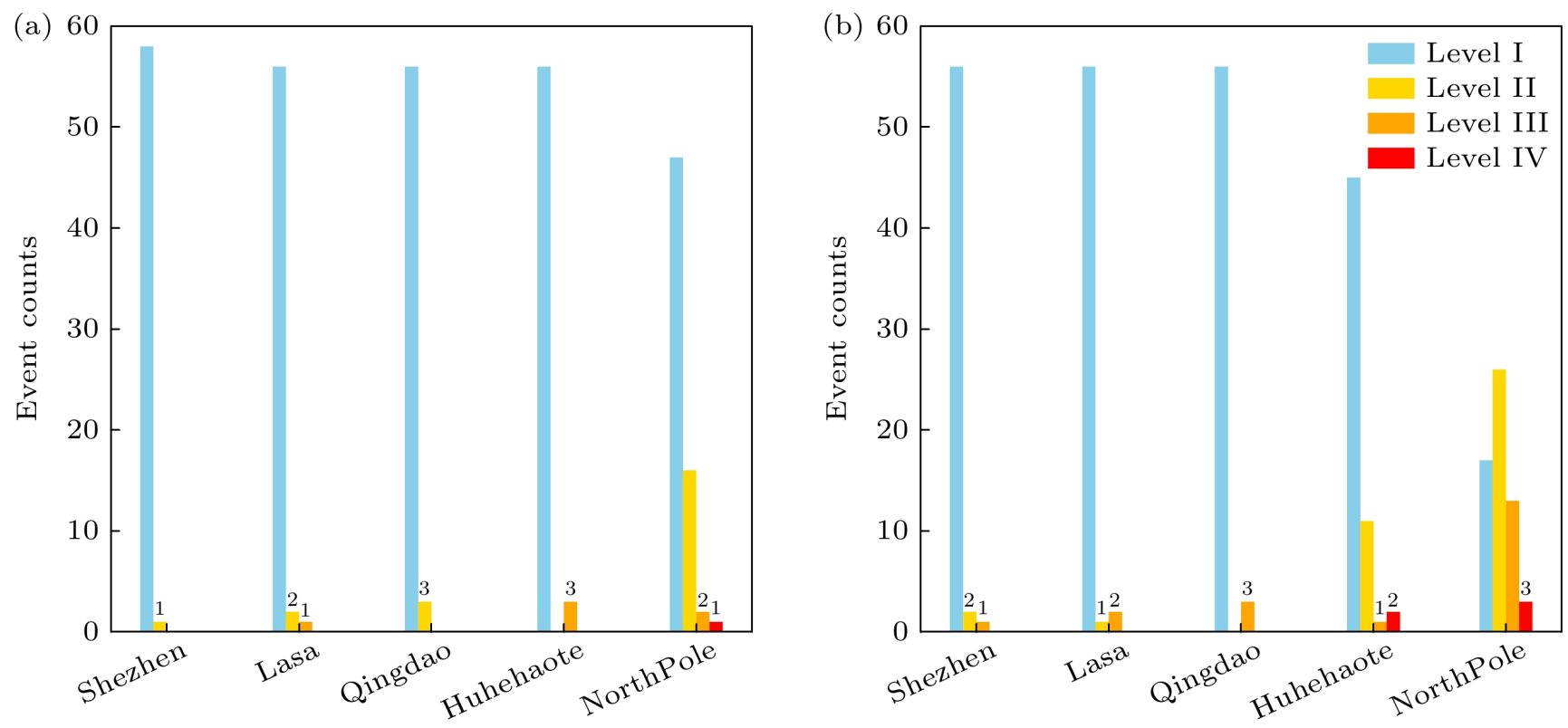

**Fig. 9 Count of GLE events for different cities from 1956 to 2012: (a) The case of isotropic injection; (b) the case of extreme anisotropic injection.**

For low-latitude cities such as Shenzhen, under isotropic injection conditions, all GLE events except the Level II event in 1956 resulted in Level I muon hazard levels. This indicates that GLE events do not exacerbate the risk of muon-induced single-event effects (SEE). Even under extreme anisotropic injection conditions, only two Level II events and one Level III event occurred over the 68-year period. Therefore, the muon SEE risk posed by GLE events is negligible.

For mid-latitude cities, Lhasa and Qingdao experienced a total of only two or three Level III events over the 68-year period, even considering extreme anisotropic injection. Thus, the muon SEE risk from GLE events is negligible, and the high altitude of Lhasa does not significantly increase this risk. In contrast, Hohhot, which is at a higher latitude with a smaller cutoff rigidity ($R_c$), experienced two Level IV events over the 68 years under extreme anisotropic injection. Consequently, increased muon SEE risk during GLE events requires attention in this region.

The situation in high-latitude regions, such as the North Pole, is less favorable. Under extreme anisotropic injection conditions, a Level III or higher event occurs approximately every three years, with frequent occurrences during solar active periods. Therefore, precautions against increased muon SEE risk during GLE events are necessary in high-latitude areas.

Since most cities in China are located in low and mid-latitude regions, the muon SEE risk to low-altitude aircraft (with 1 MB system memory) from recorded GLE events (integration time 1800 s) is negligible for most Chinese cities, with the exception of a few high-latitude locations.

# 4 Discussion

This risk assessment relies on the electronic simulation work of Hubert et al.[4] to estimate muon single-event cross-sections for different technologies and processes. However, Figure 5 clearly shows significant uncertainty in the cross-sections reported by different experiments and simulations, which complicates further quantitative assessment. Furthermore, recent muon irradiation experiments indicate that negative muons undergo capture reactions in electronic devices, producing protons and $\alpha$ particles and depositing charge. This results in a larger single-event cross-section compared to positive muons[30,31]. The differing behaviors of positive and negative muons introduce substantial uncertainty into cross-section calculations. Therefore, more precise quantitative assessments require support from additional irradiation experiments and electronic simulations.

Currently, only Sato et al.[14] have provided the energy spectrum of low-energy muons (20-400 MeV) using a full-absorption muon energy spectrometer (FAMES). Their measurements at 100 MeV are approximately 50% higher than the simulation results presented in this study. This discrepancy may stem from proton and neutron backgrounds that were not effectively subtracted in the experiment, requiring further experimental verification. If this discrepancy is considered an additional source of uncertainty, the muon SEE risk assessment in this study may be somewhat underestimated. Future implementation of more measured experiments on low-energy muon spectra will effectively reduce current uncertainties.

The preceding discussion evaluated the muon SEE risk from GLE events only under a fixed integration time window. However, it should be noted that secondary particles in GLE events arrive at the ground within a short time window and may exhibit high anisotropy in the early stages of the event[34]. This implies that real-world conditions may more closely resemble the results given by extreme anisotropic injection. Moreover, the muon SEE risk during the time window of concentrated secondary particle injection may be higher than estimated. Additionally, Navia et al.[16] reported an enhancement in muon flux associated with solar flare events, without observing an increase in neutron flux or significant flare intensity. If this phenomenon is genuine, new models must be established to assess the risk of muon SEE during solar activities such as flares.

# 5 Conclusion

This study systematically evaluated the risk of muon-induced single-event effects (SEE) for low-altitude aircraft in different cities and under various solar activity scenarios. The main conclusions are as follows.

Using the CORSIKA software package with a localized atmospheric model and precise primary particle spectra, we obtained atmospheric muon energy spectra in the MeV-GeV range for different urban areas. The reliability of these spectra was verified by comparison with BESS experimental data.

Under static conditions with only cosmic ray injection, the low-altitude environment in all Chinese cities faces significant muon SEE risk for flight control systems (1 MB memory) using advanced process ($\leqslant$ 45nm) bulk technology chips, particularly in high-altitude regions like Lhasa. In contrast, flight control systems using FD-SOI technology chips do not face significant risk.

Under static conditions, for large memory systems (1 GB), the MCU event rate for small-process chips exceeds the requirements of the highest automotive functional safety level (ASIL-D), even when using the most radiation-resistant FD-SOI technology. Therefore, hardening measures such as redundancy must be implemented.

We proposed a classification method for muon hazard levels during GLE events and assessed the muon SEE risk in different regions during these events. The assessment indicates that for most low and mid-latitude cities in China, the muon SEE risk from historical GLE events is negligible. However, in high-latitude regions, strong GLE events may lead to a significant increase in risk, warranting caution.

# Appendix A

Table A1. Altitude, longitude, latitude, and geomagnetic field parameters of different cities.

| City (Position) | Altitude | Geomagnetic Field | | Rigidity |
|---|---|---|---|---|
| | $H$/m | $B_x$/μG | $B_z$/μG | $\phi$/GV |
| Shenzhen (Lat.22°40′N, Lon.113°12′E) | 100 | 37.7 | 26.0 | 15.36 |
| Qingdao (Lat.36°12′N, Lon.120°15′E) | 50 | 30.5 | 42.4 | 9.49 |
| Hohhot (Lat.40°49′N, Lon.111°39′E) | 1050 | 27.2 | 49.0 | 6.71 |
| Lhasa (Lat.29°36′N, Lon.91°06′E) | 3650 | 34.4 | 36.8 | 12.49 |
| North Pole (Lat.90°00′N, Lon.0°0′E) | 0 | 1.8 | 56.9 | 0 |
| Lynn Lake (Lat.56°30′N, Lon.101°0′W) | 360 | 10.4 | 59.4 | 0.27 |
| Mount Norikura (Lat.36°06′N, Lon.137°33′E) | 2770 | 30.3 | 36.1 | 10.54 |
| Tsukuba (Lat.36°12′N, Lon.140°06′E) | 30 | 30.0 | 35.5 | 9.07 |

Table A2. The number of simulated primary cosmic rays in each energy range, where $n$ is the nucleon number of the corresponding particle.

| Particles | $E_{tot}/n$ /(GeV • $n^{-1}$) | | | | | |
|---|---|---|---|---|---|---|
| | 0.938—$10^1$ | $10^1$—$10^2$ | $10^2$—$10^3$ | $10^3$—$10^4$ | $10^4$—$10^5$ | $10^5$—$10^6$ |
| H | $4 \times 10^7$ | $8 \times 10^6$ | $5 \times 10^5$ | $1 \times 10^5$ | $2 \times 10^4$ | $1 \times 10^3$ |
| He | $2 \times 10^7$ | $1 \times 10^6$ | $1 \times 10^5$ | $2 \times 10^4$ | $4 \times 10^3$ | $1 \times 10^3$ |
| C | $3 \times 10^5$ | $5 \times 10^4$ | $1 \times 10^4$ | $2 \times 10^3$ | $1 \times 10^3$ | $1 \times 10^3$ |
| O | $3 \times 10^5$ | $5 \times 10^4$ | $1 \times 10^4$ | $2 \times 10^3$ | $1 \times 10^3$ | $1 \times 10^3$ |
| Fe | $1 \times 10^5$ | $1 \times 10^4$ | $3 \times 10^3$ | $1 \times 10^3$ | $1 \times 10^3$ | $1 \times 10^3$ |